\documentclass[%
 aip,
 rsi,%
 amsmath,amssymb,
 reprint,%
]{revtex4-1}

\usepackage{graphicx}
\usepackage{dcolumn}
\usepackage{bm}

\begin{document}

\title[Dry demagnetization cryostat for sub-millikelvin helium experiments]
{Dry demagnetization cryostat for sub-millikelvin helium experiments: refrigeration and thermometry}

\author{I.\,Todoshchenko} 
\address{O.\,V.\,Lounasmaa Laboratory, Aalto University, 00076~AALTO, Finland}
\email{todo@boojum.hut.fi}
\author{J.-P.\,Kaikkonen}
\address{O.\,V.\,Lounasmaa Laboratory, Aalto University, 00076~AALTO, Finland}
\author{R.\,Blaauwgeers}
\address{BlueFors Cryogenics Ltd, Arinatie 10, 00370 Helsinki, Finland}
\author{P.\,J.\,Hakonen}
\address{O.\,V.\,Lounasmaa Laboratory, Aalto University, 00076~AALTO, Finland}
\author{A.\,Savin}
\address{O.\,V.\,Lounasmaa Laboratory, Aalto University, 00076~AALTO, Finland}

\begin{abstract}
We demonstrate successful "dry" refrigeration of quantum fluids down to $T=0.16$\,mK by using copper
nuclear demagnetization stage that is pre-cooled by a pulse-tube-based dilution refrigerator. This type
of refrigeration delivers a flexible and simple sub-mK solution to a variety of needs including
experiments with superfluid $^3$He. Our central design principle was to eliminate relative vibrations
between the high-field magnet and the nuclear refrigeration stage, which resulted in the minimum heat
leak of $Q=4.4$\,nW obtained in field of 35\,mT.

For thermometry, we employed a quartz tuning fork immersed into liquid $^3$He. We show that the fork
oscillator can be considered as self-calibrating in superfluid $^3$He at the crossover point from
hydrodynamic into ballistic quasiparticle regime.
\end{abstract}

\maketitle

\section{Introduction}
The currently ongoing conversion of sub-Kelvin refrigerators to cryogen-free platforms \cite{Uhlig02} 
facilitates new applications (microscopy, imaging, medicine, space applications, security, astronomy, etc) 
as such systems can be run practically anywhere. Nowadays, dry dilution refrigerators equipped with large 
superconducting magnets and base temperatures below 10\,mK are offered by many suppliers as standard products.
Many demanding quantum experiments require a further reduction of temperature. 

To meet the above demand, the next challenge is to extend the operation of dry systems to sub-millikelvin
temperatures. Nuclear demagnetization cooling \cite{Lounasmaa} combined with  a commercially available dry
dilution refrigerator as a precooling systems is one of the options to reach the microKelvin regime. 
Recently, successful operation of such a cryogen-free experimental platform down to 600\,$\mu$K was
demonstrated by Batey {\it et al.}~\cite{Saunders2013} using a PrNi$_5$ nuclear stage. However, this
temperature is the practical limit for PrNi$_5$ because of its large intrinsic field, in contrast to
copper which enables much lower temperatures.  Additionally, copper is more available, easier to handle, and
it provides better thermal conductivity when compared with PrNi$_5$. As a consequence, most of the modern
sub-mK nuclear demagnetization refrigerators employ Cu for the nuclear cooling stage 
\cite{Berglund72,Bradley84,Huiki86,Borovik-Romanov87,Gloss88,Xu92}. However, Cu stage is
more sensitive to external heat loads because of the very demanding precooling conditions to fully polarize
the nuclear spins in copper. Furthermore, due to the high electrical conductivity, a good-quality Cu nuclear
stage is extremely prone to eddy current heating. All these reasons make nuclear cooling with copper much
more difficult to implement on a dry system, but, if successful, also much more rewarding.

Nuclear demagnetization cryostat is quite complicated and laborious machine which includes typically a large
liquid helium bath, a dilution unit, and a large-bore 8-9 Tesla solenoid. The refrigerator requires a
large amount of work to construct and to keep operational \cite{Lounasmaa}, and furthermore, its operation
needs daily attention. Consequently, for a long time, the investigation of the superfluid $^3$He was a
prerogative of big laboratories. This situation will change once a pulse-tube-based dry dilution refrigerator
is combined efficiently with a nuclear cooling stage. Such a combination will allow lengthy sub-mK experiments 
with minimal attention by the operator of the refrigerator.

The aim of this work was to make the first demonstration of superfluid $^3$He refrigeration on a "dry" nuclear
demagnetization cryostat all the way down to 0.2\,mK. Our central design principle was to eliminate relative
vibrations between the high-field magnet and the nuclear refrigeration stage. This principle was found to
work quite well and a heat leak of $Q=4.4$\,nW was obtained at 35\,mT. The heat leak was found
to scale as $B^2$ upto $\sim100$\,mT, which is a clear sign of vibrational heating due to eddy currents. For
thermometry, we employed a quartz tuning fork immersed into liquid $^3$He, which indicated 0.16 mK for the
lowest temperature. Furthermore, we show that such fork oscillator can be considered as self-calibrating
in superfluid $^3$He at the crossover point from hydrodynamic into ballistic quasiparticle regime.

The paper is organized as follows. In the next section the technical details of the cryostat are given.
In Section 3 we discuss the thermometry with the tuning fork. The performance of the cryostat is described
in Section 4 where we describe the cooling cycle and give the values for the lowest temperatures 
achieved, the heat leaks in different fields, thermal conductivity of the heat switch, and other details. 
In Conclusions we summarize the results of our experiments and their analysis.

\section{The cryogen-free demagnetization refrigerator}
Our cryostat is based on the commercially available BF-LD400 dry dilution cryostat from BlueFors
Cryogenics \cite{BlueFors}. It has a two-stage pulse tube refrigerator with a base temperature of 3\,K for 
the second stage. For condensation of $^3$He-$^4$He mixture, the system employs a 2\,bar compressor, which 
can be switched off during continuous circulation. The dilution unit cools down to 7\,mK and provides 
550\,$\mu$W of cooling power at 100\,mK. The cryostat has a set of radiation shields thermally anchored at
60\,K, 3\,K, 0.7\,K ($^3$He evaporator) and at the mixing chamber temperature.  The cryostat is equipped
with a 9\,T magnet from American Magnetics which is thermally anchored to the 2nd stage of the pulse tube 
cooler. In our BF-LD400 the still pumping line and the pulse tube mounting were fitted with the damper systems
provided by Bluefors Ltd so that the pulse tube and the turbo pump were mechanically well decoupled
from the top flange of the cryostat.

\begin{figure}[t]
\centering
\includegraphics[width=0.95\linewidth]{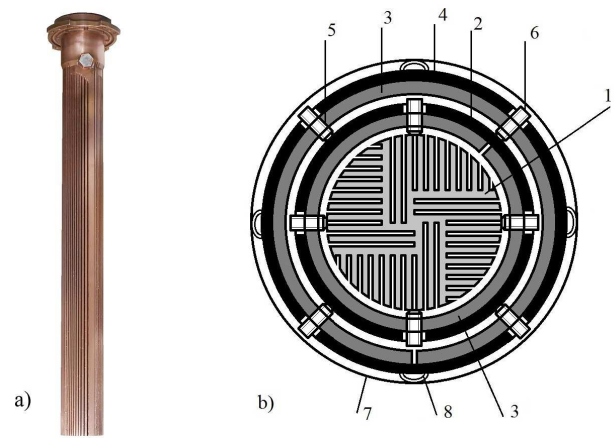}
\caption{\label{fig:spacer} a) Photograph of the copper nuclear stage with a total length of 438\,mm.
b) Cross-sectional view of the nucelear stage and radiation shields with spacers: 1 --  nuclear stage,
2 -- mixing chamber radiation shield, 3 -- slotted ring, 4 -- 0.7\,K radiation shield, 5 -- thread,
6 -- M4 nylon bolt, 7 -- bore of the 9\,T magnet, 8 -- plastic spring.
}
\end{figure}

The nuclear stage of the so-called "Helsinki design" \cite{Pekola84,Berglund89} was made of a single
cylindrical copper piece in which a set of slits were machined in order to reduce eddy currents which appear
when the enclosed magnetic flux is varying, see Fig.\,\ref{fig:spacer}. The heat switch which connects the
stage and the mixing chamber consists of seven bended aluminium foils (50\,mm$\times$10\,mm$\times$0.5\,mm 
each) diffusion welded to two copper rods, out of which one is bolted to the nuclear stage and the other one 
to the mixing chamber flange. Before welding, all parts of the heat switch were annealed: copper at
900\,$^\circ$C in $2\cdot10^{-3}$\,mbar of air for two weeks and aluminium at 550\,$^\circ$C in better
than $10^{-5}$\,mbar vacuum for 1 hour. A small superconducting solenoid, surrounded by a cylindrical niobium
shield, is mounted on the heat switch. A field of 20\,mT is employed to drive the aluminium from the
superconducting state to the normal state.

The nuclear stage is first pre-cooled in a field of $B=8$\,T; during this process, the heat switch is in
the normal state and thermally connects the nuclear stage to the mixing chamber via electronic thermal
conductivity. When the stage approaches the mixing chamber temperature, the heat switch is turned to
superconducting state where most of the electrons are bound into Cooper pairs and cannot conduct heat,
so that only phonon conductivity is left which is very small at 10 - 20 mK. The magnetic field is then
slowly decreased and, in ideal adiabatic conditions, the temperature of copper nuclei is lowered
proportionally to $B$ \cite{Lounasmaa}. In reality, some amount of entropy is lost during the demagnetization
process due to heat leaks which are to be reduced well below 1\,$\mu$W for successful cooling.

Owing to gradients in the polarizing field, eddy currents will be generated in constant field if the
nuclear stage vibrates with respect to the magnet. In the case of a "dry" dilution refrigerator, the level
of mechanical vibrations is significantly higher compared with traditional liquid-He-based refrigerators.
Despite the fact that the pulse tube cooler was partly decoupled mechanically from the rest of our cryostat
by flexible links, the vibrational heat load to the nuclear stage due to eddy currents was found to be a
critical factor for the cooling power of the stage. Initially, there were no spacers between the radiation
shields, the magnet, and the nuclear stage, so that the long tail of the nuclear stage vibrated with
respect to the magnet, and the eddy currents were producing a heat load of few $\mu$W to the stage in an
8\,T field. Addition of spacers, however, reduced the eddy current heating by an order of magnitude.

The employed spacer system is illustrated in Fig.\,\ref{fig:spacer}b. We have developed an easy way to make
spacers between the shields, the magnet and the copper stage for fixing them together without significant
changes to the shields. The spacers are made of slotted brass rings with four threaded holes for nylon bolts.
The rings are placed between two shields and between the 20-mK shield and the stage. The outer shield has
four holes for bolts which tightly press the inner shield, while the ring abuts the outer one,
see Fig.\,\ref{fig:spacer}b. The outer shield has plastic springs to go tightly inside the bore of the
magnet. All parts of the stage-shields-magnet assembly are thus strongly mechanically fixed together so that
they vibrate as a single piece without significant movements with respect to one another.

The experimental cell for liquid $^3$He is embedded inside the top part of the nuclear stage in a form of
a cylinder, and silver sinter with an effective area of about 20\,m$^2$ is baked on the cell walls. In this
first experiment, only an oscillating fork was installed inside the cell. Approximately 0.6\,mole of $^3$He
of 200\,ppm $^4$He purity was condensed to the cell. This resulted in a partially filled chamber with a free
liquid-vapor interface.

\section{Thermometry}
The thermometry of helium sample in the sub-millikelvin range is quite a difficult task.
The thermal boundary resistance, Kapitza resistance, between (dielectric) helium and the copper
refrigerant increases as $1/T^3$. This means that even tiny heat leak to helium sample will saturate
temperature of helium at the level determined by the heat leak and surface area of thermal contact, while
the copper is much colder. Because of this thermal decoupling of helium from the environment it is
absolutely necessary to measure the temperature of liquid $^3$He directly. There are three practical ways
to determine liquid $^3$He temperature by measuring a) nuclear magnetic susceptibility of helium with NMR
b) melting pressure with a sensitive {\it in situ} transducer, and c) density of quasiparticles with a 
mechanical oscillator. Unfortunately, the temperature dependence of susceptibility of superfluid $^3$He saturates 
below 0.5\,$T_c$, and thus NMR thermometry is not useful at the lowest temperatures. Moreover, the sensitivity 
of the melting curve thermometer also rapidly decreases at low temperatures. In contrast to these two 
thermometers, the sensitivity of a mechanical oscillator rapidly increases with lowering temperature 
\cite{Lancaster_Nature83}. In the ballistic regime below $\sim0.3\,T_c$, where the mean free path of 
quasiparticles becomes larger than the size of the oscillator, the fluid-induced damping decreases 
exponentially, $\Delta f\propto\exp{(-\Delta/T)}$ where $\Delta f$ is the width of the resonance of the 
oscillator and $\Delta$ is the superfluid energy gap. This behavior has been predicted theoretically by 
Gu$\rm\acute{e}$nault {\it et al.}~\cite{Lancaster_JLTP86} and demonstrated experimentally  by Todoshchenko 
{\it et al.}~\cite{Todoshchenko02} using a vibrating wire in combination with melting curve thermometry.

\begin{figure}[t]
\centering
\includegraphics[width=0.95\linewidth]{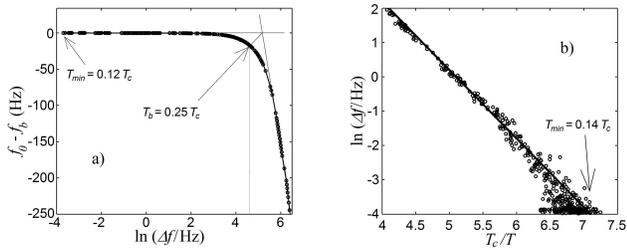}
\caption{\label{fig:ballistic_mc} Preliminary experiment on the self-calibrated fork thermometry in superfluid 
$^3$He at the melting pressure: a) Crossover from the hydrodynamic to the ballistic regime of oscillating fork 
at $\ln{(\Delta f/{\rm Hz})}\approx4.5$. The lowest temperature is 0.12\,$T_c$, according to 
Eq.\,(\ref{eq:scaling}).
b) Logarithm of the width of the resonance as a function of inverse reduced temperature measured using
melting curve thermometer. The fit shows slope -2, in agreement with Eq.\,(\ref{eq:scaling}) where
$\Delta=2.0\,T_c$ \cite{Todoshchenko02}. The lowest temperature is 0.14\,$T_c$.
}
\end{figure}

Mechanical oscillator thermometer is, however, a secondary thermometer and as such it needs to be calibrated 
against some other thermometer. Even in the ballistic regime, it needs calibration at least at one temperature 
in order to scale the width of the resonance with respect to it. Fortunately, the oscillator can be used to 
calibrate itself. The idea is to use the central frequency of the resonance as an independent single-point 
(``fixed point") thermometer. In the hydrodynamic regime there is an additional mass attached effectively to 
the oscillator due to the viscous motion of quasiparticle excitations around it.
Upon cooling, the mean free path of quasiparticles increases rapidly, and less and less quasiparticles around
the oscillator feel its motion and the effective mass of the oscillator decreases. The central frequency
of the resonance is inversely proportional to the square root of the effective mass, and it saturates
once the mean free path becomes longer than the size of the oscillator. Hence, the point of the crossover
from the hydrodynamic regime into the ballistic regime is manifested by a saturation of the dependence of
central frequency $f_0$ as a function of the width of the resonance $\Delta f$ 
(see Fig.\,\ref{fig:ballistic_mc}a).

The onset of the ballistic regime is quite exactly determined because the dependence of the mean free path
on temperature is very steep \cite{Ono82}. The calibration procedure is thus to measure the width $\Delta f_b$
at which central frequency $f_0$ saturates and attribute this width to certain temperature $T_b$
depending on the size of the oscillator. Then the temperatures below $T_b$ can be calculated according to an
exponential scaling law

\begin{equation}
\frac{\Delta f}{\Delta f_b}=\exp{[\frac{\Delta}{T_b}-\frac{\Delta}{T}]},
\label{eq:scaling}
\end{equation}

\noindent
where $\Delta f_b$ refers to the width at the onset temperature $T_b$. The superfluid energy gap $\Delta$
is temperature-independent below 0.5\,$T_c$ \cite{Muhlschlegel59} and varies with pressure from
1.8\,$T_c$ at zero bar to 2.0\,$T_c$ at the melting pressure \cite{Todoshchenko02}.
This simple exponential law together with the evident single-point calibration at the onset of the
ballistic regime makes the mechanical oscillator as a very sensitive "primary" thermometer.

The self-calibration property has proven to work reliably at the melting pressure according to independent
simultaneous temperature measurements using melting curve thermometry; this experiment was done on a regular
nuclear demagnetization cryostat for $^3$He \cite{Manninen14}. The melting curve thermometer was a
capacitive Straty-Adams strain pressure gauge \cite{Straty-Adams} with $\sim5\,\mu$bar accuracy. The resonator
in these cross-check experiments was a commercially available quartz tuning fork with 0.6\,mm wide tines,
which had exactly the same dimentions as in the present work.
According to the calculations by Ono {\it et al.}~\cite{Ono82}, the onset of the ballistic regime for this
size is at $T_b=0.25\,T_c$. The crossover from hydrodynamic to ballistic behavior is viewed best by plotting
the central frequency $f_0$ versus logarithm of the resonance width $\Delta f$ as shown in
Fig.\,\ref{fig:ballistic_mc}a. The onset of the ballistic regime, which we determine as the maximum of the second
derivative of the function $f_0=f_0(\ln{(\Delta f/{\rm Hz})})$, occurs at $\ln{(\Delta f_b/{\rm Hz})}=4.5$.
The lowest measured width corresponds to $\ln{(\Delta f_{min}/{\rm Hz})}=-3.9$ which, according to
Eq.\,(\ref{eq:scaling}), yields $T_{min}=\Delta/(4.5+3.9+\Delta/T_b)$. By substituting $\Delta=2.0\,T_c$
for the superfluid energy gap at high pressure and $T_b=0.25\,T_c$ for the onset we find $T_{min}=0.12\,T_c$.
The apparent arbitrariness in the determination of the onset location, {\it i.~e.}~$T_b$, does not affect
significantly the accuracy of the method due to sharpness of the crossover. Indeed, if we take 4 or 5 instead
of 4.5 for the $\ln{(\Delta f_b/{\rm Hz})}$, it will result in a very small change of the minimum temperature, 
ranging from 0.125\,$T_c$ to 0.118\,$T_c$. An independent measurement of $\ln{(\Delta f/{\rm Hz})}$ {\it vs} 
$T$ using melting curve thermometry, Fig.\,\ref{fig:ballistic_mc}b, yields 0.14\,$T_c$ for the minimum 
temperature. The agreement with the result $T_{min}=0.12\,T_c$ based on the self-calibration method is quite 
good if one considers the relatively large heat capacity of solid $^3$He which keeps the solid-liquid interface 
at a temperature slightly higher than that of the liquid.

\begin{figure}[t] 
\centering
\includegraphics[width=0.95\linewidth]{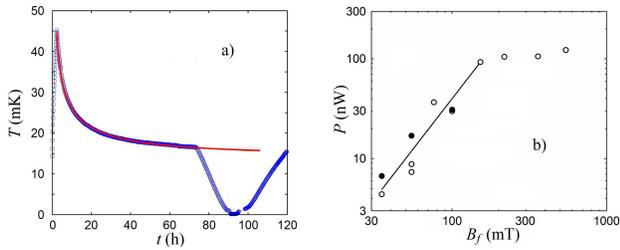}
\caption{\label{fig:heatleak} a) Temperature of liquid $^3$He inside the nuclear stage. Magnetization towards
8\,T started at time $t=0$ and was completed at $t=2.4$\,h. At $t=74$\,h the heat switch was turned off
and the magnetic field sweep from 8\,T down to 35\,mT was started. Superfluid transition was reached at time
$t=90.5$\,h in a field of 250\,mT. The curve is the fit of the precool process using Eq.\,(\ref{eq:precool}) 
with 1.4\,$\mu\Omega$ for the resistance of the heat switch and with a heat leak of 
$P=250$\,nW, see the text for detailes. 
b) The heat leak to the nuclear stage as a function of the final demagnetization field $B_f$.
Open symbols -- empty cell, closed symbols -- cell is partially filled with liquid $^3$He. The straight line
corresponds to the expected, eddy-current-induced $B^2$ dependence at low fields.}
\end{figure}

In addition to the fork resonances, the temperature of the demagnetization stage was also recorded by a noise 
thermometer manufactured by Physical-Technical Institute (Physikalisch-Technische Bundesanstalt) in Berlin 
using the electronics from Magnicon Ltd. \cite{Beyer2013} The sensitive element of the noise thermometer is a 
copper block, the thermal noise currents of which are recorded by a SQUID that is picking up the magnetic 
flux induced by the noise currents. The noise thermometer was mechanically attached to the nuclear 
demagnetization stage with 5\,mm long threaded stud at the end of the copper block. We found that this 
system worked well on the dry dilution refrigerator despite the magnetic noise from the pulse tube heat 
exchangers \cite{Eshraghi2009}. It was found that temperatures down to 0.4\,mK could be recorded on the 
nuclear stage using the noise thermometer. The saturation of the noise thermometer at temperatures below 
0.4\,mK is due to the heat leak of $\sim200$\,pW from the SQUID bias.

\section{Performance of the refrigerator}

Fig.\,\ref{fig:heatleak}a depicts a full operation cycle of our demagnetization cryostat. The cooldown starts 
by running the dilution unit at a high circulation rate of $\sim800$\,$\mu$mol$/$s, which cools the nuclear 
stage below 20\,mK through the heat switch which is kept in the conducting (normal) state by its small 
magnet. Prior to the demagnetization cycles, the experimental chamber inside the nuclear stage, was filled 
by 0.6 moles of $^3$He at low pressure to immerse the tuning fork fully inside liquid. In the normal 
fluid state above $T_c=0.93$\,mK \cite{Greywall86}, the width of the fork resonance $\Delta f$ is inversely 
proportional to temperature \cite{Eltsov2007}. The calibration of the fork for measurements in the normal state 
of $^3$He was made at $T_c$ during the demagnetization process.

\begin{figure}[t]
\centering
\includegraphics[width=0.95\linewidth]{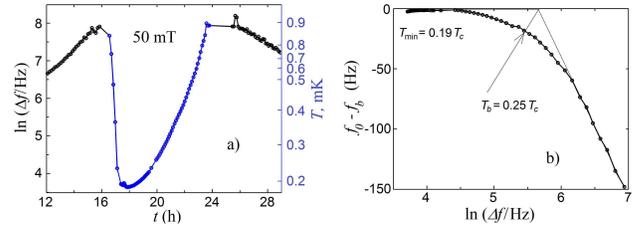}
\caption{\label{fig:ballistic} a) Logarithm of the width of the resonance in liquid $^3$He at zero pressure.
Demagnetization was started at time $t=0$ and was finished at $t=17.5$\,h. On cooling, $T_c$ was passed
at $t\simeq16.5$\,h after which the width $\Delta f$ rapidly decreased by two orders of magnitude. At 
$t = 18$\,h, warm-up began and $T_c$ was passed again at $t=24$\,h. The temperature axis on the right is 
obtained from the ballistic exponent below 0.3\,mK and using the width data by Eltsov {\it et al.} 
\cite{Eltsov2007} at 0.3\,mK~$<T<$~0.9\,mK scaled to the size of our fork. Temperature axis refers only to 
superfluid state from 16.5 to 24 hours.
b) Crossover from the hydrodynamic to the ballistic regime for the motion of the oscillating fork (from 
left to right). The saturation of the central frequency $f_0$ occurs at $\ln{(\Delta f/{\rm Hz})} \simeq 4.5$. 
The lowest temperature is 0.19\,$T_c$ according to the self-calibration method.
}
\end{figure}

In Fig.\,\ref{fig:heatleak}a, the field in the main magnet was ramped up from 0 to 8\,T in 2.5\,hours. 
Precooling of the nuclear stage at 8\,T was continued for about three days; this time is mostly determined 
by the thermal conductance of the present heat switch and the heat capacity of the nuclear stage. Our 
copper nuclear stage has effectively 3.5\,kg in the high field region, which means that its nuclear heat 
capacity is $C\simeq 1.7\cdot10^{-4}B^2/T^2$\,JK$/$T$^2=A/T^2$ \cite{Lounasmaa}, where field-dependent
constant $A$ equals 0.012\,JK in 8\,T field. The heat balance can be written as 

\begin{equation}
Pdt-\frac{A}{T^2}dT-\frac{T^2-T_0^2}{\gamma}dt=0
\end{equation}

\noindent
where $P$ is the heat leak to the nuclear stage, $\gamma$ is the thermal resistance of the heat switch, and
$T_0=14$\,mK is the ultimate temperature of the dilution unit under 8\,T conditions. After integration, we 
find

\begin{equation}
\label{eq:precool}
t-t_1=\frac{\gamma A}{\Theta^2}~[~\frac{1}{T_1}-\frac{1}{T}+
\frac{1}{2\Theta}\ln{\frac{(T_1-\Theta)}{(T_1+\Theta)}\frac{(T+\Theta)}{(T-\Theta)}}~],
\end{equation}

\noindent
where $\Theta=\sqrt{P\gamma+T_0^2}$ is a characteristic temperature of the demagnetization refrigerator, 
which accounts for the temperature of the dilution unit, for the heat leak to the nuclear stage, and for 
the quality of the heat switch. 

Eq.\,(\ref{eq:precool}) fits well the experimentally measured temperature during precooling as seen in 
Fig.\,\ref{fig:heatleak}a. From the fit we find the heat leak to the nuclear stage $P=250$\,nW and the 
thermal resistance of the switch $\gamma=120$\,K$^2/$W. If the temperature difference $\delta T=T-T_0$ over 
the heat switch is relatively small, then the heat flux through the switch can be written as 
$\dot{Q}=(2T/\gamma)\delta T$. It is more convenient to express the thermal resistance in units of 
electrical (temperature-independent) resistance $R=\gamma L/2=1.4\,\mu\Omega$, where 
$L=2.4\cdot10^{-8}$\,W$\Omega/$K$^2$ is the Lorenz number. After the precool, at time $t = 75$\,h, 
the heat switch is turned to superconducting state and the field is slowly swept down to 50\,mT.

We applied fork thermometry with self-calibration (see Sect.\,3) to measure the lowest temperature 
achieved in superfluid $^3$He on our refrigerator. Similar fork as in the experiment at the melting pressure 
was employed meaning that the onset temperature $T_b$ for the ballistic regime must again be 0.25\,$T_c$. 
Fig.\,\ref{fig:ballistic}a displays the logarithm of the fork resonance width $\Delta f$ as a function of 
time. Near $T_c$, $\Delta f$ becomes of the order of the central frequency $f_0$ and the resonance is
basically lost until the liquid enters the superfluid state at $t = 16.5$\,h. With growing superfluid 
fraction, the width decreases rapidly over two orders of magnitude and, finally at $\Delta f\sim300$\,Hz, 
the fork is in the regime where Eq.\,(\ref{eq:scaling}) becomes applicable. The liquid still cools a bit 
further down to $\Delta f=40$\,Hz which corresponds  to $T=0.19$\,$T_c=0.17$\,mK according to 
Eq.\,(\ref{eq:scaling}). While warming, $T_c$ is passed again at $t=24.0$\,h. From the warmup rate we 
can calculate the heat leak to the nuclear stage to be about 25\,nW in a final demagnetization field 
$B_f$ of 50\,mT. During the next demagnetization cycle to a field of 35\,mT, we have measured even a smaller 
fork resonance width of 15\,Hz corresponding to $T=0.16$\,mK. Note that the sensitivity of the fork thermometer
is so high at lowest temperatures that the decrease of the width by 2.5 times means only 5\,\% difference in
temperature.

During these demagnetization experiments, we also followed the nuclear stage temperature by noise thermometry. 
Unfortunately, the reading from the noise thermometer was limited to $T > 0.40$\,mK due to  overheating of 
the copper block by the SQUID bias current. At $T > 1$\,mK, the noise thermometer worked well and was 
instrumental in our heat leak measurements at high $T$.

At low values of the final demagnetization fields $B_f$, the heat leak to the nuclear stage was found to 
depend linearly on $B_f^2$ as illustrated in Fig.\,\ref{fig:heatleak}b. This means that the warmup rate 
does not actually depend on the field as both the heat capacity of the stage and the heat leak are 
proportional to $B_f^2$ (provided that $B_f \gg B_{int}=0.3$\,mT). At fields higher than 50-100\,mT, however, 
the field dependence of the heat leak becomes gradually slower and the warmup rate decreases correspondingly. 
Indeed, instead of a typical 7 hours below $T_c$, we stayed in our best cool downs at  $B_f = 100$\,mT in 
the superfluid state for 14 hours. This time span, which can be extended to $\sim 20-25$ hours by further 
demagnetization from 100 mT, can be regarded as fully adequate for basic experiments in superfluid $^3$He.
The minimum heat leak to the nuclear stage of 4.4\,nW was measured in 35\,mT field with empty cell. Typically,
the heat leaks with the cell filled with $^3$He were somewhat higher compared to empty cell because of the
viscous heating of $^3$He due to vibrations.

\section{Conclusions}

We have demonstrated  cooling of $^3$He well below the superfluid transition temperature $T=0.93$ mK 
at saturated vapor pressure using a helium-free "dry" demagnetization refrigerator. Our work is the 
first to cool helium-3 down to 0.16\,mK on a LHe-bath-free refrigerator, and thus we open an opportunity 
for deep sub-mK investigations on a commercially available pulse tube refrigerator straightforwardly 
equipped with a copper adiabatic nuclear demagnetization stage. In our work, we have introduced a new 
simple method to measure the temperature of superfluid $^3$He via the properties of mechanical fork 
oscillators without the need for other thermometers for calibration. The method has proven to work
accurately at the melting pressure when compared against an independent melting curve thermometer.

\section*{Acknowledgements}
We are grateful to Juha Tuoriniemi, Matti Manninen, and Juho Rysti who took part in the thermometry
experiments at the melting pressure. The work was supported by the Academy of Finland (contracts no.\,135908 
and 250280, LTQ CoE and FIRI2010) and the EU 7th Framework Programme (FP7/2007–2013, grant Microkelvin). 
This research project made use of the Aalto University Cryohall infrastructure.


\end{document}